%% file: AlphaLoadingTwoChamberPaper.tex
\documentclass[%
 reprint,
 amsmath,amssymb,
aip,
]{revtex4-2}

\usepackage{graphicx}
\usepackage{dcolumn}
\usepackage{bm}
\usepackage{xcolor}

\newcommand{\pa}[2]{\frac{\partial #1}{\partial #2}}
\newcommand{\paf}[2]{\partial #1/\partial #2}
\newcommand{\ve}[1]{\mathbf{#1}}

\graphicspath{{figs}}

\begin{document}


\title{Preventing ash from poisoning proton-boron 11 fusion plasmas}

\author{Ian E. Ochs}
\email{iochs@princeton.edu}
\author{Elijah J. Kolmes}
\author{Nathaniel J. Fisch}%
\affiliation{Department of Astrophysical Sciences, Princeton University, Princeton, NJ 08540}%

\date{\today}

\begin{abstract}
Proton-Boron 11 (pB11) fusion is safe and clean, but also difficult to harness for breakeven power production.
Particularly deleterious are fusion-born alpha particles, which massively increase both plasma pressure and bremsstrahlung losses unless they are pulled promptly from the plasma.
We show that even if one cannot extract the alphas quickly, one can still achieve net power production, by separating the plasma into two regions: a fusion region, accessible to all species, and an alpha storage region, accessible only to alphas and electrons.
This new demixing strategy could make pB11 fusion much easier to achieve.
\end{abstract}

\maketitle


\section{Introduction} 
Due to its abundant and safe reactants and byproducts and lack of neutron production, proton-Boron 11 (pB11) fusion has always been a theoretically appealing fusion fuel.
However, for a long time, net energy production from pB11 was dismissed as impossible, \cite{Rider1995FundamentalLimitations,Rider1995GeneralCritique} partly due to erroneously low cross section data. \cite{Nevins2000ThermonuclearFusion}
Newer cross section data \cite{Sikora2016NewEvaluation} has opened up a broader range of feasibility for both pB11 ignition \cite{Putvinski2019FusionReactivity,Ochs2022ImprovingFeasibility,Kolmes2022WavesupportedHybrid} and net energy production \cite{Ochs2024LoweringReactor}.
Simultaneously, there has been an explosion of both public and private sector interest in pB11 fusion. \cite{Rostoker1997CollidingBeam,Lampe1998NRLMR,Volosov2006AneutronicFusion,Volosov2011ProblemsACT,Labaune2013FusionReactions,Eliezer2016AvalancheProtonboron,Magee2019DirectObservation,Eliezer2020NovelFusion,Eliezer2020MitigationStopping,Ruhl2022NonthermalLaserdriven,Istokskaia2023MultiMeVAlpha,Wei2023ProtonBoronFusion,Magee2023FirstMeasurements,Liu2024ENNsRoadmap,Xie2025PreliminaryConsiderations,Liu2025FeasibilityProton}

The pB11 reaction produces 3 $\alpha$ particles, which initially contain the 8.7 MeV of energy released by the fusion reaction.
Most existing power balance analyses for steady-state fusion schemes assume that this energy quickly thermalizes and the $\alpha$'s are extracted from the plasma on a timescale much shorter than the bulk energy confinement time. \cite{Putvinski2019FusionReactivity, Ochs2022ImprovingFeasibility,Kolmes2022WavesupportedHybrid,Ochs2024LoweringReactor}
However, this may be difficult to arrange.
Thus, one must ask: what is the consequence if the $\alpha$'s cannot be quickly extracted?

As we show here, if the $\alpha$'s linger in the plasma for a time $\tau_\alpha$ equal to the energy confinement time $\tau_E$, then there are two major deleterious effects.
First, the confined pressure of the reactor increases, increasing the triple product well beyond what would be expected from an analysis excluding the $\alpha$'s.
Second, and more disastrously, the bremsstrahlung of the plasma increases to dwarf the fusion power, precluding net electrical power production by the reactor.
Both effects are serious enough to make pB11 fusion likely unworkable even if other obstacles are surmounted.
Thus, it is of paramount importance to extract the $\alpha$'s on a timescale longer than the $\alpha$-ion thermalization timescale $\tau_{\alpha i}$, but much shorter than the energy confinement time $\tau_E$.

However, it turns out that there is another solution.
The above analysis assumes a \emph{well-mixed, homogeneous} plasma.
What if, instead, one arranged a plasma with two regions: one that contained the proton and boron, and another that only the $\alpha$'s had access to? 
This would decrease the loading density of the $\alpha$ particles in the fusion region, and thus the bremsstrahlung, even at longer $\tau_\alpha \sim \tau_E$. 
Thus, such a scheme can make reactor breakeven possible, even if one cannot preferentially extract $\alpha$ particles.
Strikingly, this separation scheme can produce better results than single-region fusion even if the fuel is somewhat de-mixed as a result, departing from the conventional wisdom that a well-mixed plasma generally leads to better performance.
We provide an example showing how such desirable separation might be achieved through a combination of centrifugal and ponderomotive forces.

\section{Alpha poisoning}
The pB11 fusion reaction can be simply modeled by a set of coupled rate equations describing the change in particle density $n_\alpha$ of the $\alpha$'s, and the change in the energy density $U_s \equiv \tfrac{3}{2} n_s T_s$ of $\alpha$'s, ions, and electrons: 
\begin{align}
	\frac{dn_\alpha}{dt} &= 3 \frac{P_F}{\mathcal{E}_F} -  \frac{n_\alpha}{\tau_\alpha} \label{eq:dnadt}\\
	\frac{dU_\alpha}{dt} &= P_F +   \sum_{s \neq \alpha} K_{\alpha s} (T_s - T_\alpha) -  \frac{3}{2} \frac{n_\alpha T_\alpha}{\tau_\alpha} \label{eq:dUadt}\\
	\frac{dU_i}{dt} &=  P_H + \sum_{s \neq i} K_{i s}  (T_s - T_i) - \frac{3}{2} \frac{n_i T_i }{\tau_E} \label{eq:dUidt}\\
	\frac{dU_e}{dt} &= - P_B + \sum_{s \neq e} K_{e s}  (T_s - T_e) - \frac{3}{2} \frac{n_e T_e }{\tau_E}. \label{eq:dUedt}
\end{align}
We also assume quasineutrality, i.e. $n_e = \sum_j n_{j}$, for $j \in \{p,b,\alpha\}$.
Here, $P_H$ is the external heating power, and $\mathcal{E}_F = 8.7$ MeV is the energy released in the fusion reaction.
The $K_{i j}$ are rate constants of energy transfer collisions between species $i$ and $j$, related to the thermalization collision frequencies $\nu_{ij}$ by $K_{i j} = \frac{3}{2} \nu_{ij} n_i$, which is symmetric in $i$ and $j$ as $\nu_{ij} \propto n_j$.
$P_F$ and $P_B$ are the fusion and bremsstrahlung power densities from Refs.~\cite{Ochs2022ImprovingFeasibility,Ochs2024ErratumImproving}.
They roughly scale as:
\begin{align}
	P_F = F(T_i)  n_p n_b ; \quad P_B = B(T_e) Z_{\text{eff}} n_{e}^2, \label{eq:PfPb}
\end{align}
where $n_p$ and $n_b$ are the proton and boron densities, given as $n_p \equiv f_p n_i$, $n_b \equiv f_b n_i$, with $f_p$ and $f_b$ the proton and boron fractions of the fuel, and $f_p + f_b = 1$.
The effective charge number $Z_{\text{eff}}=\sum_j n_{j} Z_j^2/n_e$, for $j \in \{p,b,\alpha\}$.

The two characteristic timescales $\tau_\alpha$ and $\tau_E$ will be very important to the subsequent analysis.
The $\alpha$ confinement time $\tau_\alpha$ corresponds to the rate at which $\alpha$ particles are lost from the plasma, taking with them their characteristic thermal energy $3T_\alpha / 2$.
In contrast, the energy confinement time $\tau_E$ represents the characteristic timescale on which energy is lost from electrons and ions due to all non-bremsstrahlung processes.
In general, this includes both particle losses and additional radiative losses, such as from synchrotron radiation. \cite{Mlodik2023SensitivitySynchrotron,Ochs2024ElectronTail}
It is often useful to lump the total non-bremsstrahlung power losses from the plasma into a generic power loss density $P_L$: \cite{Ochs2022ImprovingFeasibility,Kolmes2022WavesupportedHybrid,Ochs2024LoweringReactor}
\begin{align}
	P_L &\equiv \frac{U_i + U_e}{\tau_E} + \frac{U_\alpha}{\tau_\alpha}.
\end{align}

Eqs.~(\ref{eq:dUidt}-\ref{eq:dUedt}) represent a simplified version of the power balance in Ref.~\onlinecite{Ochs2024LoweringReactor}, condensing protons and boron into a single population as in Refs.~\onlinecite{Putvinski2019FusionReactivity,Kolmes2022WavesupportedHybrid} and ignoring fuel burnup.
However, in contrast to those analyses, we retain the $\alpha$ density self-consistently, as was done for deuterium-tritium plasmas by Refs.~\onlinecite{Guazzotto2017TokamakTwofluid,Guazzotto2019TwofluidBurningplasma}.
Approximating the $\alpha$'s as a thermal population lends simplicity, at the cost of slightly changing the fractions of power flowing to electrons vs ions, sacrificing a slight amount of quantitative accuracy.

In solving the system of equations throughout the paper, we will set $n_i = 10^{14}$ cm$^{-3}$, $f_p = 0.85$, and $T_i = 300$ keV, which are near-optimal for thermonuclear pB11. \cite{Putvinski2019FusionReactivity,Ochs2022ImprovingFeasibility,Kolmes2022WavesupportedHybrid,Ochs2024LoweringReactor}
Note that this means that the fusion power density $P_F$ is constant throughout the analysis, while the power densities for bremsstrahlung $P_B$ and loss rates $P_L$.
By imposing $\tau_E$ and $\tau_\alpha$, we will then be able to see the effects of these parameters on solving for $n_\alpha$, $T_\alpha$, $T_e$, and $P_H$, allowing us to evaluate the plasma performance.

We can see very quickly why $\alpha$ loading is such a particular problem for pB11 fusion plasmas.
Equilibrium occurs when $\paf{}{t} \rightarrow 0$, so that Eq.~(\ref{eq:dnadt}) immediately gives:
\begin{align}
	n_\alpha = 3 \frac{P_F}{\mathcal{E}_F} \tau_\alpha. \label{eq:naFromTauA}
\end{align}
The density $n_\alpha$ can be rewritten in terms of the energy confinement time and the non-bremsstrahlung losses $P_L$.
Assuming a small fraction of $\alpha$ particles so that we can ignore the $\alpha$ contributions to $P_L$ then gives:
\begin{align}
	\frac{n_\alpha}{n_i} &= \frac{9}{2} \frac{P_F}{P_L} \frac{T_i +\langle Z_j \rangle T_e }{\mathcal{E}_F} \frac{\tau_{\alpha}}{\tau_E} \sim  0.28 \frac{P_F}{P_L} \frac{\tau_{\alpha}}{\tau_E}, \label{eq:naFromTauE}
\end{align}
where $\langle Z_j \rangle \equiv \sum_j Z_j n_i / n_e = 1.6$ is the average ion charge state [note that this is not the same as $Z_\text{eff}$].
As a breakeven fusion plasma requires $P_F / P_L \gtrsim 1$, we immediately see that if $\tau_\alpha = \tau_E$, then $n_\alpha \gtrsim n_i/3$.

As a result of the excess $\alpha$ population, the plasma pressure $p$ increases. 
Assuming the $\alpha$'s are well-thermalized, the pressure will increase by an amount:
\begin{align}
	\frac{\Delta p}{p_0} = \frac{n_\alpha}{n_i}\frac{T_i + Z_\alpha T_e}{T_i + \langle Z_i \rangle n_\alpha T_e} = 1.1 \frac{n_\alpha}{n_i}.
\end{align}
Thus, we can expect an approximately 30\% increase in the plasma pressure (and thus the triple product) at the same fusion reactivity level due to the presence of the $\alpha$'s, making the reaction conditions significantly harder to achieve.

The $\alpha$ particle population also degrades performance by increasing the bremsstrahlung radiation losses.
Using Eq.~(\ref{eq:PfPb}), we can estimate the increase in bremsstrahlung losses due to the increasing $\alpha$ density:
\begin{align}
	\pa{P_B}{n_\alpha} 
	&= \frac{P_B}{n_e} \mathcal{Z}; \quad \mathcal{Z} \equiv  \frac{Z_\alpha^2}{Z_\text{eff}}  + Z_\alpha . \label{eq:dPbdna}
\end{align}
For $f_p = 0.85$, $Z_\text{eff} \approx 3$ and so $\mathcal{Z} \approx 10/3$.
In this case, the fractional increase in bremsstrahlung losses is thus:
\begin{align}
	\frac{\Delta P_B}{P_B} = \frac{n_\alpha}{n_e}\mathcal{Z}  =  \frac{n_\alpha}{n_i} \frac{\mathcal{Z}}{\langle Z \rangle} \approx 2.1 \frac{n_\alpha}{n_i}. \label{eq:DeltaPbDeltana}
\end{align}
Considering Eq.~(\ref{eq:naFromTauE}), we see that for a breakeven plasma with $\tau_\alpha \sim \tau_E$, the bremsstrahlung power will already exceed the prediction ignoring $\alpha$'s by 60\%. 
In fact, this is a slight underestimate, as it does not account for nonlinear effects as $n_\alpha$ grows larger.

\begin{figure}[t]
	\centering
	\includegraphics[width=\linewidth]{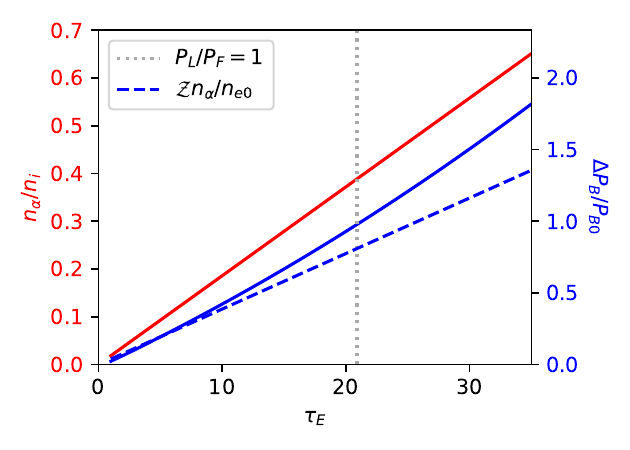}
	\caption{$\alpha$ particle density relative to fuel ions (red) and increase in bremsstrahlung power due to the presence of $\alpha$'s (blue). The gray dotted line indicates the value of $\tau_E$ at which $P_L = P_F$, and the blue dashed line represents the linearized model of bremsstrahlung in terms of $\alpha$ density [Eq.~(\ref{eq:DeltaPbDeltana})].}
	\label{fig:naAndPb}
\end{figure}

We can compare this prediction to the solution of the full system of Eqs.~(\ref{eq:dnadt}-\ref{eq:dUedt}).
We can also compare to the case without $\alpha$'s, by fixing $n_\alpha = n_i/10$ and solving Eqs.~(\ref{eq:dUadt}-\ref{eq:dUedt}) with the $\alpha$ contribution removed from the electron density, bremsstrahlung, and particle loss terms.
The solution for $n_\alpha / n_i$ and $\Delta P_B / P_{B0}$ as a function of and $\tau_E$ are shown in Fig.~\ref{fig:naAndPb}, along with the line indicating where $P_L = P_F$.
In rough agreement with (but somewhat exceeding) the simplified analysis, at this point $n_\alpha / n_i \approx 0.4$, resulting in a bremsstrahlung fraction $\approx 100$\% higher than it would be in the case without $\alpha$'s.

\section{Effect of $\alpha$'s on reactor performance}
The above simple analysis has shown that, if $n_i$ and $T_i$ are held constant by modulating the heating power $P_h$, the $\alpha$'s increase both the triple product and bremsstrahlung production of the fusion plasma.
Fundamentally, the increase in bremsstrahlung is the bigger problem of the two. 
While the triple product increase still allows net energy production, albeit at more-difficult-to-achieve plasma conditions, the bremsstrahlung increase can preclude net energy production entirely.

To see this effect on reactor performance, we calculate the engineering $Q$, \cite{Wurzel2022ProgressFusion,Ochs2022ImprovingFeasibility,Ochs2024LoweringReactor} which can be expressed as:
\begin{align}
	Q \equiv \frac{P_\text{out} - P_\text{in}}{P_\text{in}} = \eta \frac{P_L + P_B}{P_H} - 1.  
\end{align}
Here, $\eta = \eta_H \eta_L < 1$ is the product of the efficiencies $\eta_H$ of converting electrical power to heating power, and $\eta_L$ of converting bremsstrahlung and particle loss power back to electricity.
This definition of $Q$ is closely related to the electrical gain\cite{Frank2024IntegratedModelling} $Q_e \equiv P_\text{out}/P_\text{in} = Q + 1$.
Thus, it is important to remember that it is $Q > 0$ ($Q_e > 1$) that corresponds to a reactor that produces net electrical power (modulo electrical support for the confinement system).
Meanwhile, the threshold $Q > 1$ has no particular significance.

Generally, it is thought that increasing the confinement time increases the allowable $Q$.
However, taking $P_H = P_L + P_B - P_F$, and recalling that $P_F$ is constant, we find that:
\begin{align}
	\text{sgn}\left(\pa{Q}{\tau_E} \right) = 	-\text{sgn}\left( \pa{P_L}{\tau_E} + \pa{P_B}{\tau_E} \right). \label{eq:dQdTauE} 
\end{align}
In other words, increasing the energy confinement time only improves reactor performance if the resulting decrease in losses $P_L$ exceeds the increase in the bremsstrahlung $P_B$.
As we have seen, if $\tau_\alpha = \tau_E$, then increasing $\tau_E$ results in higher $n_\alpha$, which can make the second term large.

Assuming for simplicity that $T_e \approx 150$ keV is approximately constant [so that $\paf{Q}{\tau_E} = (\paf{P_B}{n_\alpha} ) \times (\paf{n_\alpha}{\tau_E})$], making use of Eqs.~(\ref{eq:naFromTauA}) and (\ref{eq:dPbdna}), and recalling that $T_i$ and $P_F$ are constant, we can readily calculate the two derivatives in Eq.~(\ref{eq:dQdTauE}):
\begin{align}
	\pa{P_L}{\tau_E} = - \frac{\sum_s U_s}{\tau_E^2}; \quad \pa{P_B}{\tau_E} = 3 \mathcal{Z} \frac{P_B  P_F}{\mathcal{E}_F n_e}. \label{eq:dPbdTauE}
\end{align}

%

\begin{figure}[t]
	\centering
	\includegraphics[width=\linewidth]{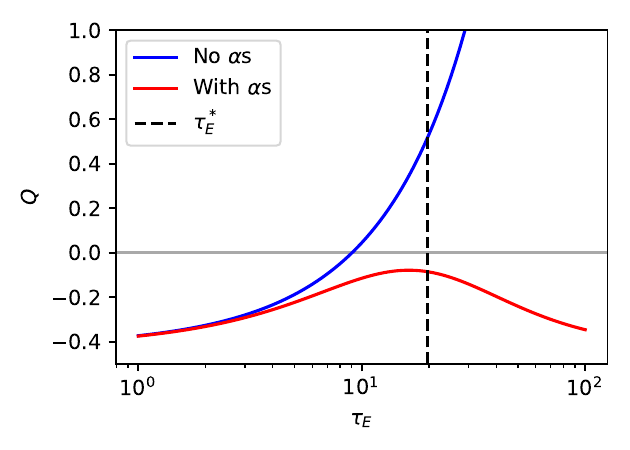}
	\caption{Reactor performance parameter $Q$ versus energy confinement time $\tau_E$ for $\eta = 0.576$.
	Two scenarios are shown:
	(blue) a scenario like Refs.~\onlinecite{Putvinski2019FusionReactivity,Ochs2022ImprovingFeasibility,Kolmes2022WavesupportedHybrid,Ochs2024LoweringReactor}, where $\alpha$'s instantly thermalize and are extracted from the plasma, so that $n_\alpha = 0$. extraction, i.e. $\tau_\alpha = 0$; vs. 
	(red) a scenario with self-consistently thermalizing $\alpha$ particles, which are extracted on a timescale $\tau_\alpha = \tau_E$.
	In the latter case, bremsstrahlung losses due to increasing $\alpha$ poisoning cause decreasing performance past $\tau_E^* \approx 20$ seconds [Eq.~(\ref{eq:tauEStar})], and reactor breakeven becomes impossible.}
	\label{fig:QVsTauE}
\end{figure}

Using Eqs.~(\ref{eq:dQdTauE}-\ref{eq:dPbdTauE}) and again ignoring $\alpha$ contributions to $P_L$, we can linearize around $P_{B0} \sim P_F$ and $n_{e0} = n_i \sum_i f_i Z_i$ to find the point $\tau_E^*$ at which $\paf{Q}{\tau_E} = 0$, i.e. where increasing $\tau_E$ no longer improves reactor performance:
\begin{align}
	\tau_E^* &= \left( 2 \mathcal{Z}  \frac{P_{B0}}{\sum_s n_s T_s} \frac{P_F}{\mathcal{E}_F n_e} \right)^{-1/2} \approx 20 \text{ s.} \label{eq:tauEStar}
\end{align}
This point is unfortunately close to the $\alpha$-free breakeven point of $\tau_E \sim 10$s, \cite{Ochs2024LoweringReactor} which is directly related to the fact that the bremsstrahlung starts to spike as $P_L \rightarrow P_F$.
Thus, even an efficient reactor with $\eta = 0.576$, representing $64\%$ efficient electrical conversion and $90\%$ efficient heating power delivery, fails to produce net power once $\alpha$ poisoning is included self consistently.
This failure to breakeven can be seen in Fig.~\ref{fig:QVsTauE}, which shows the $Q$ that results from the full solutions to Eqs.~(\ref{eq:dnadt}-\ref{eq:dUedt}) with $\tau_\alpha = \tau_E$, both with and without $\alpha$ poisoning.

\section{Selective Ash Deconfinement}
Of course, the above calculations assumed $\tau_\alpha = \tau_E$.
If we relax this requirement, we can expect to do much better; indeed, $\tau_\alpha / \tau_E$ was found to be a critical parameter in DT fusion power balances. \cite{Guazzotto2017TokamakTwofluid,Guazzotto2019TwofluidBurningplasma} 
Optimizing $Q$ over $\tau_E$ in  this case gives $\paf{Q}{\tau_E} > 0$, since $\paf{P_B}{\tau_E} \approx 0$ when $\tau_\alpha$ and $\tau_E$ are treated as independent.

It is worth asking, then, if we had perfect control over $\alpha$ extraction, how long we would want to keep the $\alpha$'s around.
Thus, examine $\paf{Q}{\tau_\alpha}$, via Eq.~(\ref{eq:dQdTauE}) with $\tau_\alpha$ in place of $\tau_E$.
While $\paf{P_B}{\tau_\alpha}$ remains the same as $\paf{P_B}{\tau_E}$ before, it is clear that $\paf{P_L}{\tau_\alpha} \ll \paf{P_L}{\tau_E}$, since now the $\alpha$ contribution to $P_L$, which we neglected before, is now the only contribution.
Furthermore, since from Eq.~(\ref{eq:naFromTauA}) $n_\alpha \propto \tau_\alpha$, the change in power loss depends only on the change in $\alpha$ temperature $T_\alpha$:
\begin{align}
	\pa{P_L}{\tau_\alpha} &= \pa{}{\tau_\alpha} \left(\frac{3}{2} \frac{n_\alpha T_\alpha}{\tau_\alpha} \right) = \frac{9}{2} \frac{P_F}{\mathcal{E}_F}  \pa{T_\alpha}{\tau_\alpha}. \label{eq:dPldTauA}
\end{align}
Combining Eqs.~(\ref{eq:dnadt}-\ref{eq:dUadt}), ignoring $\alpha$-$e$ collisions, and for simplicity taking $\nu_{\alpha i}$ independent of $T_\alpha$, we find:
\begin{align}
	T_\alpha &= \frac{T_{\alpha 0} + \tau_\alpha \nu_{\alpha i} T_i }{1 + \tau_\alpha\nu_{\alpha i}}; \quad \pa{T_\alpha}{\tau_\alpha} = -\frac{\nu_{\alpha i} (T_{\alpha 0} - T_i)}{(1 + \tau_\alpha \nu_{\alpha i})^2} . \label{eq:dTadTauA}
\end{align}
Here, $T_{\alpha 0} \equiv 2 \mathcal{E}_F / 9$ is the effective temperature of the fusion-born $\alpha$ particles.
Then, combining Eqs.~(\ref{eq:dQdTauE}), (\ref{eq:dPbdTauE}), (\ref{eq:dPldTauA}), and (\ref{eq:dTadTauA}) gives:
\begin{align}
	\tau_\alpha^* \nu_{\alpha i} &= \left(\sqrt{\frac{3}{2 \mathcal{Z}} \frac{\nu_{\alpha i}(T_{\alpha0} - T_i)}{P_{B0}/n_e}} - 1\right). \label{eq:tauAStar}
\end{align}
Note that $\nu_{\alpha i}$ is a function of $T_\alpha$, so this should be solved iteratively.
Taking $T_\alpha \sim T_i$ gives $\tau_\alpha^* \nu_{\alpha i}  \approx 8$, implying $T_\alpha \approx 500$ keV.
Plugging this into Eq.~(\ref{eq:tauAStar}), we find $\tau_{\alpha} \approx 2.6$ seconds.  
Carrying this iteraction until it converges leads to $\tau_{\alpha} \approx 2.4$ seconds.
Note that this result is independent of $\tau_E$.

\begin{figure}[t]
	\centering
	\includegraphics[width=\linewidth]{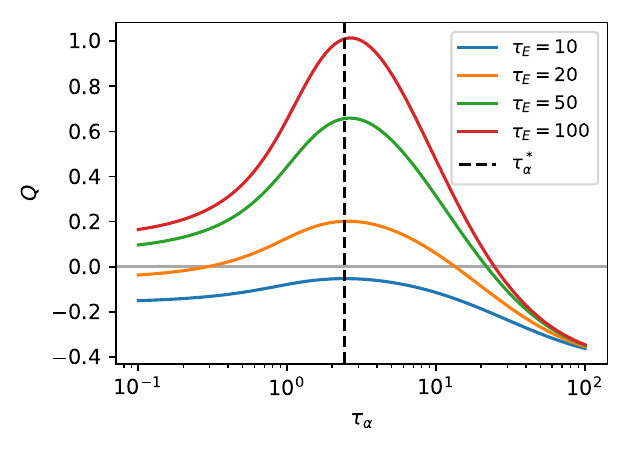}
	\caption{Reactor performance $Q$ vs. $\alpha$ confinement time $\tau_\alpha$ for several values of the energy confinement time $\tau_E$. The optimal value $\tau_\alpha^*$ [Eq.~(\ref{eq:tauAStar})] is independent of $\tau_E$, and corresponds to the point at which the upside from $\alpha$ heating of ions no longer justifies the bremsstrahlung cost of increased $\alpha$ density.}
	\label{fig:QVsTauAlpha}
\end{figure}

The physics in Eq.~(\ref{eq:tauAStar}) is clear; in the numerator, we see the initial $\alpha$-ion thermalization rate (per $\alpha$), and in the denominator, the initial bremsstrahlung rate (per electron). 
Thus, $\tau_\alpha^*$ represents a competition between increasing power delivered to ions, vs increasing power lost in bremmstrahlung.
In Fig.~\ref{fig:QVsTauAlpha}, which shows the $Q$ that results from the full solutions to Eqs.~(\ref{eq:dnadt}-\ref{eq:dUedt}) as we vary $\tau_\alpha$ at fixed $\tau_E$, we can see that Eq.~(\ref{eq:tauAStar}) accurately predicts the turnover point where holding on the the $\alpha$'s no longer improves reactor efficiency.
We also see that reactor breakeven requires $\sim 10$x separation between energy confinement time and $\alpha$ particle confinement time.

It is interesting to note that the constancy of $\tau_\alpha^*$ with $\tau_E$ means that there is a consistent optimal $\alpha$ particle density:
\begin{align}
	\frac{n_\alpha^*}{n_i} = 3 \frac{P_F \tau_\alpha^*}{n_i \mathcal{E}_F} \sim 4.5\%. \label{eq:nAlphaStar}
\end{align}
Evidently, this is the density that optimizes between thermalization and bremsstrahlung.
Interestingly, as we begin to look at other methods to reduce bremsstrahlung by splitting the plasma, this optimal density will remain fairly consistent.

\section{The natural synergy of pB11 fusion and $\alpha$ channeling}

Getting an $\alpha$ particle confinement time that is lower than the typical energy confinement time doesn't seem like it should be that difficult; after all, even in many DT fusion schemes, $\alpha$ particles are promptly lost due to their comparatively large orbits at MeV-scale energies, in addition to fast-ion-generated instabilities.
However, the above analysis has come with a caveat; the $\alpha$ particles must be removed, but \emph{only once they have transferred their energy to the bulk plasma}.
Otherwise, the reactor cannot achieve optimal performance (though this result can be mitigated by a sufficiently efficient direct conversion scheme.\cite{Ochs2024LoweringReactor,Volosov2006AneutronicFusion,Volosov2011ProblemsACT})

The necessity of allowing the $\alpha$'s to thermalize creates a problem: once thermalized, $\alpha$'s are much less distinguishable from the other particles.
Their charge-to-mass ratio is similar to boron, and their speed is not much different from the fast protons. 
Thus, any wave-based scheme which seeks to target thermal $\alpha$ particles is likely to dump out large amounts of proton and boron as well.

These problems are solved if a wave can be put in the plasma which interacts with the high-energy $\alpha$ particles, simultaneously extracting the $\alpha$ particles from the plasma while harvesting their energy.
If this energy can be transferred into a wave that heats the fuel ions, then one has solved all the problems.
Namely, one has (i) harvested all $\alpha$ particle energy to drive further fusion, and (ii) extracted the $\alpha$ particles, reducing bremsstrahlung.
Identifying such waves is the basis of the theory of $\alpha$ channeling, whether in tokamaks, \cite{Fisch1992CurrentDrive,Fisch1992InteractionEnergetic,Valeo1994ExcitationLargekTheta,Fisch1995AlphaPowera,Herrmann1997CoolingEnergetic,Ochs2015AlphaChanneling} mirrors, \cite{Fisch2006AlphaChanneling} or rotating mirrors. \cite{Fetterman2008AlphaChanneling}
While the utility of $\alpha$ channeling has been recognized mainly for improving the reactivity of DT fusion plasmas,\cite{Fisch1994UtilityExtracting} its advantage turns out to be even greater for the pB11 reaction, in which $\alpha$ poisoning and bremssstrahlung play a much larger role.

\section{Splitting the plasma}
While wave-based diffusion provides a possible method for reducing $\tau_\alpha$, it might not always be so easy to arrange for the correct waves.
Thus, it makes sense to ask if there are any other ways to modify the plasma in order to reduce the $\alpha$ density.

Thus, consider a plasma with various potentials present (centrifugal, electrostatic, ponderomotive).
Using these potentials, it is possible to create regions of the plasma that favor the presence of one species or another.
Then, if one creates a second, tenuous but large region that only $\alpha$'s and electrons can access, it will have a similar effect to reducing $\tau_\alpha$: it will reduce the $\alpha$ density in the fusion region (and thus the bremsstrahlung power) without significantly increasing other loss terms.

To see this in action, consider a two-region plasma with a fusion region $F$ and an $\alpha$ storage region $H$, with volumes $V_F$ and $V_H$ respectively.
The total number of particles of species $s$ is then $N_s = n_{s}^F V_F + n_{s}^H V_H$, where the superscipt indicates the region where the measurement is taken.

For more direct comparability to our previous equations, define $\bar{N}_s \equiv N_s/V_F$; this reduces to $n_{s}^F$ when $V_H = 0$.
Then, for any species $s$:
\begin{align}
	\bar{N}_s &= n_s^F \left(1+ \Lambda_s \right); \, \bar{U}_s = U_s^F \left(1+ \Lambda_s \right); \, \Lambda_s \equiv \bar{V} \bar{n}_{s}^H.
\end{align}
where we have defined $\bar{V} \equiv V_H / V_F$ and $\bar{n}_{s}^H \equiv n_{s}^H/n_{s}^F $.
Note that $\Lambda_s$ is the ratio of the total number of particles in chamber $H$ to chamber $F$ for species $s$.
Then, summing Eqs.~(\ref{eq:dnadt}-\ref{eq:dUedt}) over both regions (while assuming a single temperature for each species), we find
a set of equations of the same form as Eqs.~(\ref{eq:dnadt}-\ref{eq:dUedt}), with the substitutions $n_s \rightarrow \bar{N}_s$, $U_s \rightarrow \bar{U}_s$, $K_{ij} \rightarrow \bar{K}_{ij}$, and $P_B \rightarrow \bar{P}_B$, where:
\begin{align}
	\bar{K}_{ij} & \equiv K^F_{ij} + K^H_{ij} \approx K^F_{ij} \left(1 + \bar{V}\bar{n}_{i}^H \bar{n}_{j}^H \right)\\
	\bar{P}_B &\equiv P_{B}^F \left(1+  \bar{V} \bar{n}_{eH}^2 \frac{Z_{\text{eff}}^H}{Z_{\text{eff}}^F}  \right),
\end{align}
and where we must now satisfy quasineutrality separately in each chamber.

We take the limit of a large and tenuous $H$ region via: 
\begin{align}
	\bar{n}_{s}^H = \delta \sqrt{\Lambda_s}; \; \bar{V} =  \sqrt{\Lambda_s}/\delta; \; \delta \rightarrow 0.
\end{align}
In this limit, collisions and bremsstrahlung in chamber $H$ become negligible, i.e. $\bar{K}_{ij} \rightarrow K^F_{ij}$ and $\bar{P}_B \rightarrow P_B^F$.

Now, only allow $\alpha$'s and electrons access this tenuous region $H$ (i.e. take $\Lambda_p,\Lambda_b \rightarrow 0)$, and take $\tau_\alpha = \tau_E$.
Then the power balance is entirely determined by $\tau_E$ and $\Lambda_\alpha$, with $\bar{N}_e$ and $\Lambda_e$ determined by the requirements of quasineutrality.
For a given $n_i^F$ and $\tau_\alpha$, $\bar{N}_s$ will be independent of $\Lambda_\alpha$ for all $s$, since the $\alpha$ population is given by Eq.~(\ref{eq:naFromTauA}) with $n_\alpha \rightarrow \bar{N}_\alpha$, i.e.:
\begin{align}
	\bar{N}_\alpha = 3 \frac{P_F}{\mathcal{E}_F} \tau_\alpha. \label{eq:NaBarFromTauA}
\end{align}
As a result, the ion and electron loss terms will not increase as a result of including the extra volume.
Meanwhile, the bremsstrahlung power (which is negligible in chamber $H$) will be substantially reduced.

\begin{figure}[t]
	\centering
	\includegraphics[width=\linewidth]{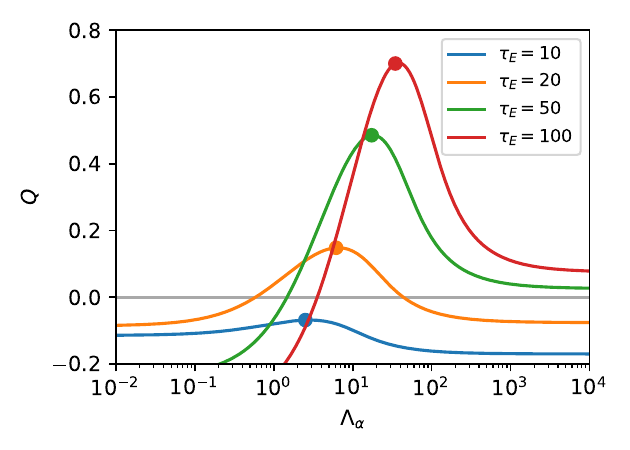}
	\caption{Reactor performance $Q$ versus $\alpha$ fraction in second region $\Lambda_\alpha$ for several values of the energy confinement time $\tau_E$, for $\tau_\alpha = \tau_E$. 
		The optimal value $\Lambda_\alpha^*$ [Eq.~(\ref{eq:LambdaAlphaStar})]   is proportional to $\tau_\alpha$, and is represented by the large dots.
		Similarly to $\tau_\alpha^*$ in the one-chamber case, $\Lambda_\alpha^*$ corresponds to the point at which the upside from $\alpha$ heating of ions no longer justifies the increased bremsstrahlung from high $\alpha$ density in the fusion region.}
	\label{fig:QVsLambdaAlpha}
\end{figure}

To see this explicitly, we can calculate $\paf{P_L}{\Lambda_\alpha}$ and $\paf{P_B}{\Lambda_\alpha}$.
It is straightforward to see that increasing $\Lambda_\alpha$ reduces the bremsstrahlung losses.
From Eq.~(\ref{eq:dPbdna}):
\begin{align}
	\pa{P_B}{\Lambda_\alpha} &= \pa{P_B^F}{n_\alpha^F} \pa{n_\alpha^F}{\Lambda_\alpha} = - \mathcal{Z} \frac{P_B^F}{n_e^F} \frac{\bar{N}_\alpha}{(1+\Lambda_\alpha)^2}.  \label{eq:dPbdLambda}
\end{align}
Because $\bar{N}_s$ is independent of $\Lambda_\alpha$, increasing $\Lambda_\alpha$ (like decreasing $\tau_\alpha$) increases $P_L$ only mildly, due only to the increased $\alpha$ temperature from the reduced thermalization rate $\bar{K}_{\alpha i} \approx K_{\alpha i}/(1+\Lambda_\alpha)$, with $T_\alpha$ thus given by:
\begin{align}
	T_\alpha &= \frac{T_{\alpha 0}(1 + \Lambda_\alpha) + \tau_\alpha \nu_{\alpha i}^F T_i }{(1 + \Lambda_\alpha)  + \tau_\alpha\nu_{\alpha i}^F}.
\end{align}
Taking the derivative with respect to $\Lambda_\alpha$ yields:
\begin{align}
	\pa{P_L}{\Lambda_\alpha} &= \pa{P_L}{T_\alpha} \pa{T_\alpha}{\Lambda_\alpha} =  \frac{3}{2}   \frac{\bar{N}_\alpha \nu_{\alpha i}^F (T_{\alpha 0} - T_i)}{(1 + \tau_\alpha \nu_{\alpha i}^F + \Lambda_\alpha)^2}  \label{eq:dPldLambda}
\end{align}

Taking the relevant limit $\tau_\alpha \nu_{\alpha i}^F \gg (1,\Lambda_\alpha)$, we can combine Eqs.~(\ref{eq:dPbdLambda}) and (\ref{eq:dPldLambda}) to find $\Lambda_\alpha^*$ where $\paf{Q}{\Lambda_\alpha} = 0$:
\begin{align}
	\frac{\Lambda_\alpha^* + 1 }{\tau_\alpha \nu_{\alpha i}^F} &= \frac{1}{1+\tau_\alpha^* \nu_{\alpha i}^F}, \label{eq:LambdaAlphaStar}
\end{align}
where $\tau_\alpha^*$ comes from Eq.~(\ref{eq:tauAStar}), with $\nu_{\alpha i} \rightarrow \nu_{\alpha i}^F$.

The fact that $(1+\Lambda_\alpha)$ is proportional to $\tau_\alpha = \tau_E$ means that, once again, the optimal configuration is characterized by a single optimal fusion-chamber $\alpha$ density, regardless of $\tau_E$. Using Eq.~(\ref{eq:NaBarFromTauA}):
\begin{align}
	n_{\alpha}^{F*} = \frac{\bar{N}_\alpha}{1 + \Lambda_\alpha^*} 
	= n_\alpha^* \left(1 + \frac{1}{\tau_\alpha^* \nu_{\alpha i}^F} \right),
\end{align}
where $n_\alpha^*$ is given by Eq.~(\ref{eq:nAlphaStar}).
Because $\tau_\alpha^* \nu_{\alpha i}^F \gg 1$, whether we are extracting the $\alpha$'s or thinning them out using a separate chamber, the optimal density of $\alpha$'s in the fusion region stays approximately constant, increasing only slightly to $n_\alpha^{F*} / n_i^F = 5.2\%$.

The $Q$ that results from the full solution of the two-region power balance equations, with $\tau_\alpha = \tau_E$, $\bar{n}_\alpha^H = 1/10$ and $\bar{V} = 10 \Lambda_\alpha$, are shown in Fig.~\ref{fig:QVsLambdaAlpha}.
Roughly, this plot appears as a scaled mirror image of Fig.~\ref{fig:QVsTauAlpha}.
Small $\Lambda_\alpha$ corresponds to taking the limit $\tau_\alpha \sim \tau_E$, precluding breakeven (as in Fig.~\ref{fig:QVsTauE}), while large $\Lambda_\alpha$ corresponds to the limit $\tau_\alpha \ll  \tau_E$, where the $\alpha$'s are low density, but are dumped before they can transfer significant energy to the ions. 
The intermediate value of $\Lambda_\alpha^*$, like $\tau_\alpha^*$, represents a point optimized accounting for both $\alpha$ energy transfer to ions and low bremsstrahlung losses.

\section{Additional advantages}
In addition to an increase in the reactor performance $Q$, there are also two other distinct advantages to either selectively deconfining or splitting out the $\alpha$'s.

\begin{figure}[t]
	\centering
	\includegraphics[width=\linewidth]{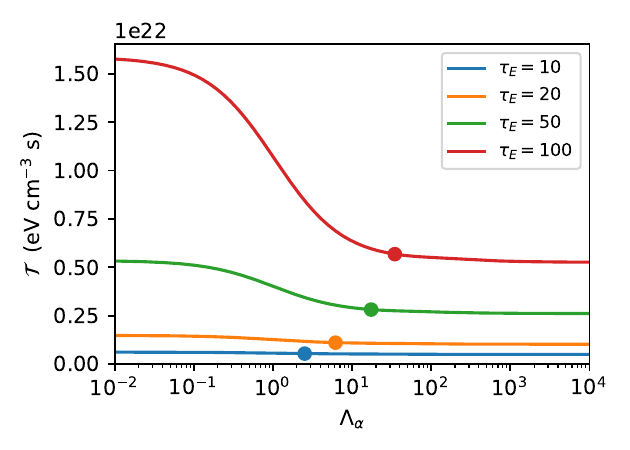}
	\caption{Triple product in fusion region $\mathcal{T} \equiv \sum_s n_s^F T_s \tau_{Es}$ vs. $\alpha$ fraction in second region $\Lambda_\alpha$ for several values of the energy confinement time $\tau_E$, for $\tau_\alpha = \tau_E$. 
	The optimal $\Lambda_\alpha$ for each $\tau_E$ is shown as a dot.
	For a given $\tau_E$, the reduction in the $\alpha$ density in the fusion region leads to a decrease in the triple product; an effect that becomes particularly pronounced at high $\tau_E$, where the loading from $\alpha$'s grows large.}
	\label{fig:TripleProductVsLambdaA}
\end{figure}

First, as discussed earlier, the difficulty of confining a fusion plasma is generally treated as a function of the triple product of the density, temperature, and energy confinement times of the plasma constituents.
Really, this should be considered a product of the plasma pressure and maximal energy confinement time:
\begin{align}
	\mathcal{T} =  \tau_{E} \sum_s n_s T_s.
\end{align}
We have seen that for $\tau_E = \tau_\alpha$, the $\alpha$ fraction becomes large quickly, contributing significantly to $\mathcal{T}$.
However, because the optimal $\alpha$ density is around $n_\alpha^F \sim 5\%$ of $n_i^F$, near the optimal values $\tau_\alpha^*$ or $\Lambda_\alpha^*$, the $\alpha$'s contribute negligibly (around 7\%) to $\mathcal{T}$.
Thus, at the same time as the deconfinement or separation techniques allow for higher plasma performance at a given $\tau_E$, they also make that $\tau_E$ easier to achieve for a given $n_i^F$ by reducing the associated triple product $\mathcal{T}$.
This reduction in $\mathcal{T}$ with increasing $\Lambda_\alpha$ is shown in Fig.~\ref{fig:TripleProductVsLambdaA} for the same set of simulations as in Fig.~\ref{fig:QVsLambdaAlpha}.
For the high-performance case of $\tau_E = 100$s, the reduction in $\mathcal{T}$ is greater than 50\%.

Second, one of the primary advantages of pB11 fusion is that it is aneutronic.
However, if high energy $\alpha$'s coexist alongside boron, a``side-chain'' fusion reaction can occur, \cite{Walker1949CrossSection,Bonner1956NeutronsGamma,Liu2020MeasurementAlpha,Hora2021EliminationSecondary} producing a nitrogen and a neutron, which receives the bulk of the kinetic energy of the reactants (up to $\sim 6$ MeV, but more typically $\lesssim 3$ MeV).

Because the $B-\alpha$ reaction cross section is a strong function of the $\alpha$ energy, with reactions coming primarily from $\gtrsim 1$ MeV $\alpha$'s, one cannot simply look at the $\alpha$ density to evaluate the reduction in side-chain reaction rate. 
Instead, one must consider the population of fast $\alpha$'s, which fundamentally requires a kinetic analysis.
Nevertheless, certain results can be intuited.

For the case of $\alpha$ deconfinement, if one waits for the $\alpha$ particles to thermalize before extracting them on a timescale $\tau_\alpha \sim \tau_\alpha^*$, then there will be no reduction in side chain reactions, as any side chain reactions which would have occurred would occur before the particle is thermalized.
Thus, extracting $\alpha$'s after collisional thermalization does nothing to reduce side-chain reactions.

There is a significant caveat to this conclusion, however.
Consider again the case of $\alpha$ channeling, where the $\alpha$'s are extracted through quasilinear diffusion by a well-chosen wave extracts energy from the $\alpha$ particles and puts it into the ion population.
This effectively raises the thermalization rate between the $\alpha$'s and the ions, allowing the $\alpha$'s to be extracted on a timescale faster that $\tau_\alpha^*$ without sacrificing the performance upside from $\alpha$-ion thermalization.
The resulting low $\alpha$ density would have the simultaneous merits of bringing reactor performance in line with $\alpha$-free analyses \cite{Putvinski2019FusionReactivity,Ochs2022ImprovingFeasibility,Kolmes2022WavesupportedHybrid,Ochs2024LoweringReactor} and reducing side chain reactions (if $\tau_\alpha \nu_{\alpha i} \lesssim 1$).
Once again, we see that $\alpha$ channeling pairs very naturally with pB11 fusion.

Now consider the case of a separated plasma.
If the separation was achieved as the result of a potential barrier, with $0 < \psi_\alpha / T_\alpha \ll (\psi_p / T_i, \psi_b/T_i)$, then targeting a relatively low density $\alpha$ population in the $H$ region will imply $\psi_\alpha$ is on the order of a few $T_\alpha$.
Thus, the most reactive $\alpha$'s--those at or above 2 MeV--will pass over the potential easily.
Very roughly, then, the reaction rate will be reduced by the ratio of the $H$ chamber volume to the $F$ chamber volume, i.e. by a factor $(\bar{V} + 1) \equiv (\Lambda_\alpha / \bar{n}_\alpha^F + 1)$.
This can be a massive factor; $\sim 180$ for the $\tau_E = 50$s case frin Fig.~\ref{fig:QVsLambdaAlpha}, and $\sim 360$ for $\tau_E = 100$.

To summarize, while both $\alpha$ extraction and separation can significantly reduce the triple product at a given fuel ion density and temperature, $\alpha$ extraction can only reduce side-chain reactions if it is combined with enhanced $\alpha$-ion thermalization with extraction occuring on a timescale shorter than the fusion timescale, while $\alpha$ separation naturally reduces side-chain reactions by a large factor.

\section{Imperfect separation}

So far, we have shown that separating out $\alpha$'s into a tenuous second plasma region can be advantageous.
However, it might not always be possible to isolate $\alpha$'s completely.
For instance, if (as proposed) access to the two regions is controlled by a potential difference $\psi_s$ for each species as it passes from $F$ to $H$, then the density of each species in each chamber will scale as:
\begin{align}
	\bar{n}_s^H = e^{-\psi_s/T_s}; \label{eq:nsBarFromPsi}
\end{align}
i.e. requiring a vanishing proton and boron population in chamber two requires extremely large, species-dependent potentials to be produced in the plasma, which can be technically difficult.
Thus, it makes sense to ask what happens if the separation is not so perfect, and some other ion species gain access to the $H$ chamber.

Ignore for now the boron, whose large mass makes it fairly easy to contain via centrifugal forces, but consider that some protons might reach the $H$ chamber.
Since in our analysis we hold $n_p^F$ fixed, the additional proton population increases $\bar{N}_p$, and thus increases proton losses.
Ambipolarity constraints ($\bar{N}_e = \sum_j Z_j \bar{N}_j$) means that it also increases $\bar{N}_e$, and thus electron losses as well.
Thus, as we increase the fraction of protons $\Lambda_p$ in the $H$ chamber, the losses change by an amount:
\begin{align}
	\pa{P_L}{\Lambda_p} = \pa{P_L}{\bar{N}_p} \pa{\bar{N}_p}{\Lambda_p} +  \pa{P_L}{\bar{N}_e} \pa{\bar{N}_e}{\Lambda_p} = \frac{3}{2} \frac{n_i^F(T_i + T_e)}{\tau_E}. 
\end{align}

\begin{figure}[t]
	\centering
	\includegraphics[width=\linewidth]{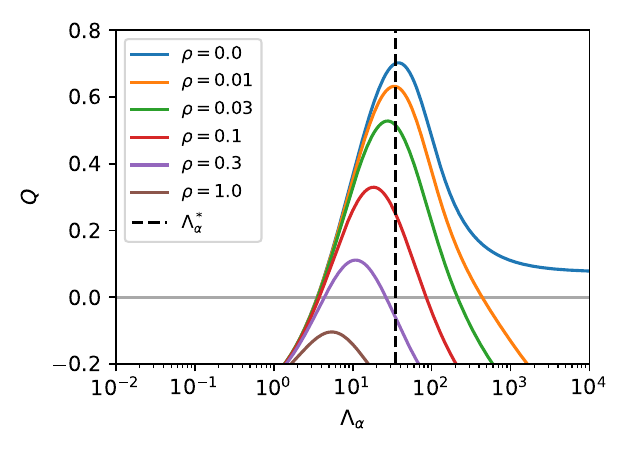}
	\caption{Reactor performance $Q$ versus $\alpha$ fraction in second region $\Lambda_\alpha$ for  $\tau_E = \tau_\alpha = 100$s, for several values of the proton-to-$\alpha$ density ratio fraction $\rho = \Lambda_p / \Lambda_\alpha$. 
		At $\rho^* \sim 0.03$, the performance begins to be significantly impacted by enhanced proton losses, though positive-$Q$ performance is possible even up to $\rho \sim 0.3$.}
	\label{fig:QVsLambdaAlphaForRho}
\end{figure}

Now we can examine the resulting impact of the $H$-region protons on reactor performance.
As a measure of proton poisoning in the $H$ chamber, define $\rho \equiv \bar{n}^H_p /\bar{n}^H_\alpha = \Lambda_p / \Lambda_\alpha$.
With $\rho$ fixed, Eq.~(\ref{eq:dPldLambda}) becomes:
\begin{align}
	\pa{P_L}{\Lambda_\alpha} &=  \frac{3}{2}  \frac{\bar{N}_\alpha \nu_{\alpha i}^F (T_{\alpha 0} - T_i)}{(1 + \tau_\alpha \nu_{\alpha i}^F + \Lambda_\alpha)^2}  + \rho \frac{3}{2} \frac{n_i^F(T_i + T_e)}{\tau_E}. \label{eq:dPldLambdaAWithProtons}
\end{align}
The performance of the two-region plasma will significantly decline relative to the proton-free case when the second term in Eq.~(\ref{eq:dPldLambdaAWithProtons}) begins to dominate.
This occurs when [using Eq.~(\ref{eq:NaBarFromTauA}) and recalling $\tau_\alpha \nu_{\alpha i}^F \gg (1,\Lambda_\alpha)$, and $T_{\alpha 0 } \gg T_i$]:
\begin{align}
	\rho^* 
	&\sim \frac{2}{3}\frac{P_F}{\nu_{\alpha i } n_i^F (T_i + T_e)} \frac{\tau_E}{\tau_\alpha}. \label{eq:rhoStar}
\end{align}
The right hand side of Eq.~(\ref{eq:rhoStar}) is a ratio of the fusion rate to a modified thermalization rate, so it is fairly small; indeed, when $\tau_\alpha = \tau_E$, we have $\rho^* = 0.03$.
We can see in Fig.~\ref{fig:QVsLambdaAlphaForRho} what happens as we repeat the simulations for $\tau_E = 100$s in Fig.~\ref{fig:QVsLambdaAlpha}, but now with $\rho \neq 0$.
Both the the optimal value of $\Lambda_\alpha$ and the performance $Q$ are reduced substantially as $\rho$ increases past $\rho^*$, with positive-$Q$ operation ceasing around $\rho \sim 0.5$.

Thus, we see that the relative fraction of protons in $H$ chamber must be substantially less than the relative fraction of $\alpha$'s in this chamber; i.e. $\bar{n}_p^H \ll \bar{n}_\alpha^H$.
Thus, both protons and boron must see a larger potential as they enter the $H$ chamber than $\alpha$'s do.

\section{Achieving Separation}

We have seen that the best advantages to pB11 fusion come when we can separate $\alpha$ particles from both proton and boron.
Inconveniently, $\alpha$ particles have both a mass and a charge that are intermediate between the proton and boron mass and charge.
Thus, one might think that it would be impossible to find a potential configuration that allows $\alpha$ particles into a certain region, while excluding proton and boron, since this requires $\psi_\alpha/T_\alpha \ll (\psi_p/T_i,\psi_b/T_i)$.
Happily, at least one solution exists, if one is willing to make use of ponderomotive potentials.

\begin{figure}[t]
	\centering
	\includegraphics[width=0.9\linewidth]{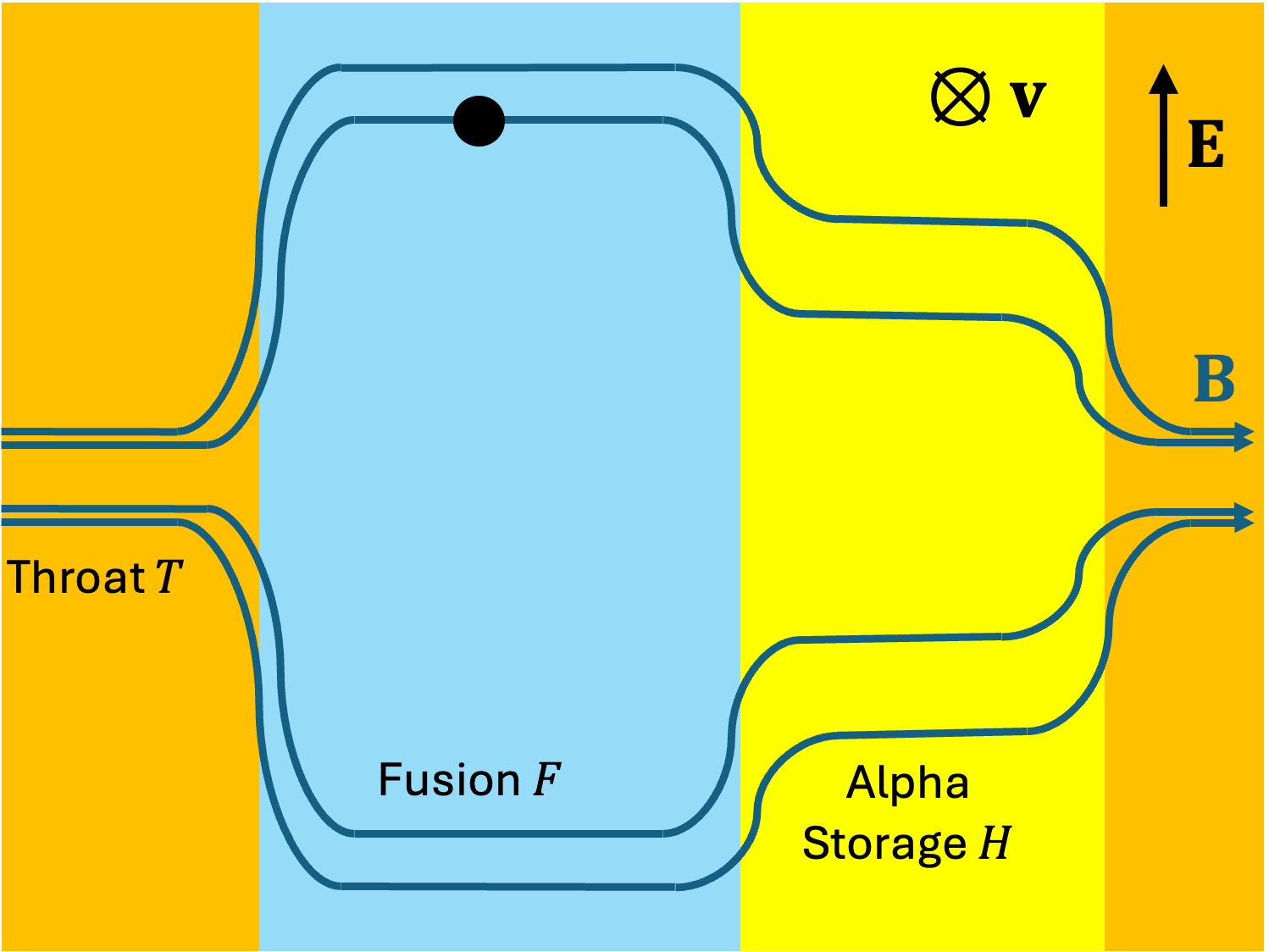}
	\caption{Schematic of a two-region centrifugal mirror, with axial magnetic field $\ve{B}$ and radial electric field $\ve{E}$ causing $\ve{E} \times \ve{B}$ rotation.}
	\label{fig:CentrifugalMirror}
\end{figure}

To begin, consider a magnetic centrifugal mirror.
Bend field lines so there is no change in field strength, but there is a change in radius, with a higher radius for region $F$ and a lower radius for region $H$ (Fig.~\ref{fig:CentrifugalMirror}).
In this configuration, going from region $F$ to region $H$, each ion species $j$ sees a change in centrifugal potential $\psi_{Cj}$, related to the proton centrifugal potential by: 
\begin{equation}
	\psi_{Cj} = \psi_{Cp} \mu_j, \label{eq:psiCentrifugal}
\end{equation}
where $\mu_j$ is the proton-normalized mass of species $j$.

To this configuration, add transversely-polarized waves near the ion cyclotron frequency with a gradually sloping envelope. 
As a result of these waves, each ion species $j$ will see a change in the ponderomotive potential $\psi_{Pj}$, given by: \cite{Gaponov1958PotentialWells,Pitaevskii1961ElectricForces,Motz1967RadioFrequencyConfinement,Dodin2004PonderomotiveBarrier,Miller2023RFPlugging}
\begin{equation}
	\psi_{Pj} = \frac{Z_j^2 e^2 | E |^2}{4 m_j (\omega^2 - \Omega_j^2)},
\end{equation} 
where $\omega$ is the wave frequency and $\Omega_j \propto Z_j /\mu_j$ is the cyclotron frequency of species $j$.
As for the centrifugal potential, we can express the $\psi_{Pj}$ in terms of the proton cyclotron frequency:
\begin{equation}
	\psi_{Pj} = \psi_{Pp} \frac{Z_j^2}{\mu_j} \frac{(1 - \bar{\Omega}_p^2)}{(1 - \bar{\Omega}_j^2)},
\end{equation}
where $\bar{\Omega}_j = \Omega_j/\omega$.

A key feature of the above ponderomotive potential is that the $\bar{\Omega}_j$ dependence allows it to take a different sign for protons than it does for $\alpha$'s and boron, since the latter have a cyclotron frequency around half the size of the former.
This ultimately allows us to build the desired potential configuration, as follows.

First, recall that $\psi_{Cj} > 0$ $\forall j$, so that the centrifugal potential repels all ion species from region $H$.
Then choose $\psi_{Pp} > 0$ and $1 < \bar{\Omega}_p < 2$, implying that $\psi_{Pj} < 0$ for $j\in\{b,\alpha\}$.
Thus, the centrifugal potential repels protons from region $H$, while attracting borons and $\alpha$'s.

Second, note that the strengths of these potentials are not the same for each species. 
In particular, for the boron, the centrifugal potential and ponderomotive potentials are both stronger than for the $\alpha$'s.
Furthermore, the ponderomotive potential is slightly proportionally weaker for the boron than for the $\alpha$'s, since:
\begin{align}
	\frac{\psi_{Pj}}{\psi_{Cj}} \propto \frac{Z_j^2}{\mu_j^2} \frac{1}{1 - \bar{\Omega}_j^2},
\end{align}
and boron as a slightly lower charge-to-mass ratio (and thus also slightly lower $\bar{\Omega}_j$) than the $\alpha$'s.
The above scalings allow us to choose a ponderomotive potential that nearly cancels the centrifugal potential for the $\alpha$'s, leading to a slight net positive potential, while leaving a deep centrifugal well for the boron.
In this way, we can make an arbitrarily deep potential well for protons and boron, confining them to region $F$, while leaving the $\alpha$'s relatively free to traverse both regions.

It is worthwhile noting also that, since the plasma is rotating, similar potentials can also be produced by static perturbations in magnetic or electric fields imposed on the periphery, which then appear as waves with finite frequency in the rotating frame of the plasma.  
This method can be technologically advantageous, using simpler engineering components and drawing less power than wave-injection methods. \cite{Rubin2023MagnetostaticPonderomotive,Rubin2023GuidingCentre,Ochs2023CriticalRole,Rubin2024FlowingPlasma,Kolmes2024CoriolisForces,Rubin2025PonderomotiveBarriers}

In the above analysis, we neglected one additional potential: the ambipolar potential.
This potential arises because electrons do not see either of the above potentials.
Thus, to enforce quasineutrality, an electric potential $\phi$ must form between the regions, resulting in each species seeing an additional potential energy:
\begin{equation}
	\psi_{Es} = Z_s \phi.
\end{equation}
To solve for this ambipolar potential, note that in terms of the $F$-chamber density $n_s^F$ of each species, the $H$-chamber density is given by:
\begin{equation}
	n_s^H = n_s^F e^{-\sum_W \psi_{Ws} / T_s}; \quad W\in\{C,P,E\}. 
\end{equation}
Taking $n_e^F = \sum_j n_j^F$, $\phi$ is then determined by enforcing ambipolarity in the $H$ region:
\begin{equation}
	\sum_s Z_s n_s^H = 0. \label{eq:Ambipolar}
\end{equation}

To determine the full equilibrium, we must therefore solve Eqs.~(\ref{eq:psiCentrifugal}-\ref{eq:Ambipolar}).
This can be done numerically fairly straightforwardly.

As an example, consider an $F$-region fuel ion density $n_i^F = 10^{14}$ cm$^{-3}$ of 85\% protons and 15\% boron, with an additional added 5\% $\alpha$ density, with temperatures $T_i = 300$ keV, $T_e = 150$ keV, and $T_\alpha = 500$ keV.
In this case, a proton centrifugal potential of $\psi_{Cp} = 1.2$ MeV and a proton ponderomotive potential of $\psi_{Pp} = 810$ keV (at $\bar{\Omega}_p = 1.5$) results in an $\alpha$ population moderately reduced to $\bar{n}_\alpha^H  = 10\%$ of the $F$ chamber value, while the proton population is reduced to $n_p^H / n_p^F = 1.1\%$ and the boron value is vanishingly small.
Thus, for this configuration, $\rho \equiv \bar{n}_p^H / \bar{n}_\alpha^H \approx 0.1$, allowing for breakeven fusion reactor operation.

\section{Discussion and Conclusion}

In the above analysis, we have established that $\alpha$ particle management is critical to achieving steady-state breakeven pB11 fusion, both to avoid the increase in the triple product, and to reduce the excess bremsstrahlung losses.
This management can take either the form of rapid $\alpha$ particle extraction, or separation of the plasma into a fusion region $F$ and an $\alpha$-sequestration region $H$.
The separation strategy has the additional benefit of naturally reducing the side-chain reaction rate, reducing deleterious neutron production.
Furthermore, we showed that $\alpha$ sequestration could be achieved through the use of a combination of centrifugal and ponderomotive potentials.

Of course, the desirability of the separation strategy depends on many other factors not considered here.
For instance, generation of the ponderomotive potentials requires the presence of large standing wave energy in the plasma.
While this energy need not necessarily dissipate in order to provide a ponderomotive potential, leakage of this energy could lead to a large loss term. 
Similarly, the use of large centrifugal potentials can lead to losses associated with dissipation of the rotational energy.
And, of course, the $H$ chamber represents a large magnetized volume of the reactor that must be supported by the confinement system, which could require large power input to maintain.

It is important to point out that while our example made use of ponderomotive potentials, it is by no means clear that these are necessary to produce the desired results.
For instance, one could envision making the $H$ region at higher potential than the $F$ region, even for $\alpha$ particles, but having an additional potential barrier to pass between the regions.
In such a case, the high-energy fusion-born $\alpha$'s could initially have access to both regions, but then fall into one region or the other as they slow down. 
One could then attempt to manipulate the diffusion rates to ensure that more $\alpha$'s fall into the $H$ region, allowing $\alpha$ sequestration without the use of wave-based potentials.
Of course, the details of such a scheme require a fundamentally kinetic analysis outside the scope of this paper.

Along with potential downsides, creating a second plasma region also produces significant possible opportunities.
For instance, we discussed above that it might be desirable to employ $\alpha$ channeling in order to extract $\alpha$ particles from the plasma while heating ions.
One of the best ways to do this might be by targeting the cyclotron harmonics; however, since boron and $\alpha$ particles have very similar cyclotron frequencies, this would be likely to primarily heat the boron, which is known to produce less fusion reactivity than heating the hot protons.
However, if a second region is present, which only $\alpha$ particles and hot protons have access to, then any wave process will necessarily transfer power between the $\alpha$'s and the hot protons, preferentially heating the most reactive part of the proton distribution and thus dramatically improving the power balance. \cite{Ochs2022ImprovingFeasibility,Kolmes2022WavesupportedHybrid,Ochs2024LoweringReactor}
Thus, it could easily turn out that an optimal configuration employs a combination of strategies, simultaneously separating out the $\alpha$'s and extracting their energy quickly with waves, to leverage multiple of the above benefits.
What is certain, however, is that any steady-state pB11 fusion reactor must have a strategy to deal with the accumulation of $\alpha$ particle ash, while still capturing its power to fuel the fusion reaction.

\section*{Acknowledgements}

This work was supported by ARPA-E Grant No. DE-AR0001554. 

\section*{Data Availability}

Data sharing is not applicable to this article as no new data were created or analyzed in this study.


\input{AlphaLoadingTwoChamberPaper.bbl}
\clearpage
\newpage


\end{document}

%% file: AlphaLoadingTwoChamberPaper.bbl
%